# A Trapped Field of >3T in Bulk MgB$_2$ Fabricated by Uniaxial Hot Pressing

J H Durrell[1], C E J Dancer[2], A Dennis[1], Y Shi[1], Z Xu[1], A M Campbell[1], N H Babu[1,3], R I Todd[2], C R M Grovenor[2] and D A Cardwell[1]

1. Department of Engineering, University of Cambridge, Trumpington Street, Cambridge, CB2 1PZ, UK.
2. Department of Materials, University of Oxford, 16 Parks Road, Oxford, OX1 3PH, UK.
3. BCAST, Brunel University, Uxbridge, Middlesex, UB8 3PH, UK

**Abstract:** A trapped field of over 3 T has been measured at 17.5 K in a magnetised stack of two disc-shaped bulk MgB$_2$ superconductors of diameter 25 mm and thickness 5.4 mm. The bulk MgB$_2$ samples were fabricated by uniaxial hot pressing, which is a readily scalable, industrial technique, to 91% of their maximum theoretical density. The macroscopic critical current density derived from the trapped field data using the Biot-Savart law is consistent with the measured local critical current density. From this we conclude that critical current density, and therefore trapped field performance, is limited by the flux pinning available in MgB$_2$, rather than by lack of connectivity. This suggests strongly that both increasing sample size and enhancing pinning through doping will allow further increases in trapped field performance of bulk MgB$_2$.

1. Introduction

Bulk MgB$_2$ has been reported to exhibit similar performance at 20 K in terms of critical current density and field trapping ability to that of the rare-earth barium cuprate [(RE)BCO] bulk superconductors at 77 K [1]. Such temperatures may be attained relatively easily via the use of cryo-cooling systems, the cost of which continues to decrease. (RE)BCO materials are challenging to process into large geometries that carry large currents, however, due to grain boundary connectivity issues [2], so it is therefore difficult to take advantage of one of the key features of superconducting quasi-permanent magnets, that trapped field scales with sample diameter, $d$ (with energy density scaling as $d^2$). This should be compared to conventional permanent magnet materials, where remanent field and energy density are fixed materials properties and do not vary with sample size.

A simple and reliable production method is fundamental to the practical fabrication of any technical material. Two successful methods reported in the literature for the fabrication of MgB$_2$ are based on Mg infiltration [3,4] and on hot pressing [5,6]. A useful review of MgB$_2$ preparation routes is given by Dancer [7], who reports that uniaxial hot pressing is as effective as the more expensive and complex hot isostatic processing technique for the processing of bulk MgB$_2$. Several authors have reported that the use of either pressure or sintering alone is not sufficient to produce well-connected MgB$_2$ bulk material. This observation is important because grain boundaries in bulk MgB$_2$ do not form a strong barrier to current flow [8,9], as is the case in (RE)BCO [10] and pnictide superconductors [11], although poor grain connectivity does affect strongly their superconducting properties [12]. Recent work by Yamamoto *et al.* suggests that, with MgB$_2$ conductor applications in mind, ex-situ processing methods produce a dense, compact grain structure to yield optimum sample properties [13]

Actual trapped fields observed in bulk $MgB_2$ have been somewhat lower than anticipated by considering the critical current densities measured locally in $MgB_2$ bulk samples. Viznichenko *et al.* [6] have reported trapped fields of 1.5 T at 20 K and over 2 T at 4.2 K in a $MgB_2$ cylinder of diameter 28 mm and height 11 mm. These samples exhibited relatively high values of critical current density, $J_c$, of the order of $2 \times 10^9$ $A/m^2$. However, both Viznichenko *et al.* [6] and Mikhenko *et al.* [14] have suggested that $J_c$ values of $7-9 \times 10^9$ $A/m^2$ and trapped fields of 4 T at 20 K and 7 T at 4.2 K are possible theoretically. Following on from earlier work by Giunchi *et al.* [3,15], Perini [16] has reported more recently a trapped field of 1 T in a bulk $MgB_2$ sample fabricated by an Mg infiltration growth technique. Hsieh *et al.* [17] further reported local $J_c$ values in zero field of up to $8 \times 10^8$ $A/m^2$ in samples fabricated by uniaxial hot pressing. Even more recently, Naito *et al.* [1] have reported fields of up to 1.5 T in $MgB_2$ bulk samples produced by a novel hot pressing technique where the pressure is generated by the expansion of the pellet itself with increasing temperature. Naito *et al.* note explicitly that fields of the order of 3 T should be achievable at ~20 K in $MgB_2$ and signpost this clearly as a community goal. Indeed, in very recent work Yamamoto *et al.* have reported 3T in samples slightly larger in diameter than those discussed here [18].

Although a target trapped field in bulk $MgB_2$ of around 3 T at 20 K appears modest compared to the record field of 17 T achieved at 29 K in (RE)BCO [19], $MgB_2$ is potentially important technologically for a number of engineering applications if it can be fabricated using a scalable, cost effective production technique.

In this article we report the field trapping performance and its relationship to local critical current density values in $MgB_2$ bulk samples fabricated by uniaxial hot pressing. We demonstrate that $MgB_2$ bulk materials can be produced in a straightforward and scalable process and, with a diameter of only 25 mm, generate trapped fields of over 3 T at 17.5 K.

2. Sample Preparation

Magnesium diboride bulk samples were fabricated from commercially available $MgB_2$ powder (Alfa Aesar, 98% purity, -325 mesh). Characterisation of this powder has been discussed in detail in a previous paper [20]. The batch of $MgB_2$ powder used for these samples has a $\theta-2\theta$ X-ray diffraction (XRD) spectrum consisting solely of peaks attributable to the $MgB_2$ phase. The hot press used in this work is homebuilt and consists of an induction coil placed within a chamber mounted on a uniaxial press. The hot pressing temperature is taken from a measurement using a thermocouple in contact with the die wall. Powders were weighed and loaded into a 25 mm diameter graphite die with samples separated by graphite foil discs and graphite spacers. Four samples were prepared simultaneously. The magnesium diboride samples were hot pressed under a flowing argon atmosphere under 25 MPa uniaxial pressure and at 1200 °C for 30 minutes. Following removal from the graphite die, the flat surfaces of all discs were ground using a flatbed grinder (Jones and Shipman) to remove the graphite foil and the carbon-contaminated surface of the samples.

The density of the samples was measured using the Archimedes method with propan-2-ol as the immersion medium. The MgO content of the samples was determined from the ratio of the areas of the $MgB_2$ (110) and MgO (220) peaks measured from X-ray diffraction spectra, using the method described previously by Dancer *et al.* [20]. The theoretical maximum density of the samples was estimated using a rule of mixtures, assuming that the samples contained only MgO and $MgB_2$. The theoretical density was compared to the measured density to obtain a better estimate of the relative density. The grain size was estimated from optical micrographs of a sample polished using diamond pastes.

## 3. Measurements

Trapped field measurements were performed by assembling two, notionally identical, MgB$_2$ bulk samples of diameter 25 mm and height 5.38 mm into a stack using STYCAST 2850FT thermally conductive epoxy resin. Five Lakeshore HGT-2101 Hall generator ICs, chosen for their small size (0.6 mm in height), were embedded in the resin in the small gap between the superconductors.

The Hall probes were mounted across the centre of the composite sample at locations of –8, –4, 0, 4 and 8 mm from the centre of the stack. An individual calibration curve for each Hall IC was measured at 40 K, having verified experimentally that the V/T characteristics of these sensors do not vary by more than 2% between 40 K and 20 K.

The assembled superconductor/Hall generator stack, shown in Fig. 1, was cooled in a magnetic field of 5 T in a superconducting magnet containing a variable temperature insert. The externally applied field was reduced at 0.03 T/min to avoid flux jumps or avalanches once the desired measurement temperature had been achieved.

The trapped field measured in this way approximates more closely the field that would be found in the centre of a long monolithic sample, where $B_t=\mu_0 J_c r$, although in the relatively thin samples measured here a more complex approach based on Biot-Savart had to be used to relate trapped field to macroscopic critical current density [1].

Specimens were cut from the centre and the edge of one parent bulk MgB$_2$ sample to measure the local critical current density and M vs H loops using a Quantum Design MPMS-XL SQUID magnetometer. The critical current density was determined from the width of the M-H loop using the extended Bean model [21]. In addition, a SQUID magnetometer was used to determine that the critical temperature of the MgB$_2$, $T_{onset,}$ was 38 K at both the edge and central regions of the sample.

## 4. Results and Discussion

The relative density of each MgB$_2$ bulk sample was measured to be 91% of the theoretical maximum value (taking account of the MgO content) and contained ~10 % crystalline MgO phase. Some small peaks attributable to MgB$_4$ were also observed in the XRD spectra. The MgO content is slightly higher than in samples produced previously using vacuum spark plasma sintering [22], which is likely to be due to the use of inert gas as a protective atmosphere rather than sintering in a vacuum. Although the samples were dense with a characteristic golden lustre following polishing, some residual porosity could be observed by optical microscopy, which correlates well with the measured bulk density. The grain size was estimated to be ~3 μm, which is again consistent with previous studies [22].

The trapped field achieved when the sample was cooled in an externally applied field of 5 T to 20 K and 17.5 K is shown in Fig. 2. An attempt was made to field cool (FC) the sample to 15 K but, even with the slow field ramp rate used, a flux avalanche resulted in the loss of flux from the sample. This suggests that the trapped field in MgB$_2$ bulk samples is susceptible to flux jumps and avalanches, as has been observed at lower temperatures [22]. The MgB$_2$ stack was observed to trap 3.14 T at 17.5 K and 2.6 T at 20 K, values which are significantly larger than those reported previously. It should be noted, however, that the MgB$_2$ stack arrangement used to measure the trapped field, as with the record field observed for YBCO by Tomita *et al.* [19], does allow larger values of trapped field to be recorded than those measured at the top surface of a single sample. The trapped field is approximately equal to the applied field for an applied field greater than about 4 T, which suggests this field represents a limit to the flux trapping ability of the bulk MgB$_2$ sample at 17.5 K.

While a relatively complex equation may be derived from the Biot-Savart law, assuming constant critical density, for the field at any height ($z$) above a superconducting sample with a certain radius ($a$) and height ($t$) [1], it is sufficient in the geometry adopted in the present study (i.e. with the Hall probes embedded between two bulk samples) to consider the field at the surface of one sample as representative of 50% of the two sample arrangement. By neglecting the gap between the samples occupied by the Hall probes, we are essentially relating critical current density to the field that would be trapped at the centre of a single sample of height $2t$. In this case the field and critical current density are therefore related by the following expression, where $B_t$ is the maximum trapped field on the sample axis:

$$B_t = J_c \mu_0 a k \qquad (1)$$

where $k$ is the correction factor to the simple Bean model due to the finite height of a disc-shaped sample given by:

$$k = \frac{t}{a} \ln\left(\frac{a}{t} + \sqrt{1 + \left(\frac{a}{t}\right)^2}\right) \qquad (2)$$

The bulk transport critical current density for the samples used in the present study was derived from equation (2) to be 3.04 x$10^8$ A/m$^2$ at 17.5 K and 2.51x$10^8$ A/m$^2$ at 20 K. These values are comparable to those reported by other authors for good quality MgB$_2$ (as discussed in the Introduction). The value of $k$ is 0.66 for the samples studied here, which indicates that there is substantial scope to increase the field trapping ability of the system by increasing the height of the superconducting samples with respect to their diameter. It should be noted that it is possible to fabricate samples with height equal to diameter by the uniaxial hot pressing technique [23].

Figure 3 shows the trapped field profile in the samples at 17.5 K and 20 K. This is approximately conical in cross-section, indicating that the sample is indeed in the critical state. Divergence from the simple Bean model is expected because of the $J_c(B)$ relationship (illustrated in Fig. 4), which tends to reduce the critical current density towards the centre of the sample.

Two specimens of approximate dimensions 2x2x5 mm$^3$ were cut from a parent MgB$_2$ sample to investigate the relationship between the local critical current density and that derived from the measured trapped field. The critical current density determined from a magnetisation loop measured at 17.5 K using the Bean model extended for the case of a rectangular cross section [21] is shown in Fig. 4.

The value of critical current density in self-field was measured locally to be 7x$10^8$ A/m$^2$ at the edge of the sample and 8x$10^8$ A/m$^2$ at its centre. Previous authors have compared the value of bulk critical current density with the value in self-field determined locally from magnetisation measurements. For example, Naito et al [1] reported local J$_c$ values of 1x$10^9$ A/m$^2$ and values of between 4 and 5x$10^8$ A/m$^2$ from trapped field measurements. This leads generally to the conclusion that the bulk critical current density is limited by sample microstructure and can reasonably be expected to increase with improvements in processing.

To provide a better comparison between local and bulk fields, therefore, finite element modelling [24] of the measurement system was performed, taking into account the actual gap between the two bulk samples of 0.6 mm, as well as the dimensions of the samples. The usual boundary condition, that the field is zero at several diameters away from the sample, was used. The measured average self-field current density of 8x$10^8$ A/m$^2$ was used in the calculations. The Kim model [25], in which $J_c =$

$J_{c0}\left(1+\frac{B}{B_0}\right)^{-1}$, was used with $B_0$=0.6 T selected to approximate the measured $J_c(B)$ behaviour. The Kim model, chosen for simplicity, begins to diverge from the experimental data at ~1.5-2 T and the fit becomes increasingly poor at higher fields. However, given that the trapped flux density is determined mostly by the high critical current density at low magnetic fields, a reasonable approximation of the field profile is still obtained from the finite element model when the $J_c(B)$ behaviour is approximated using the Kim model.

The field profile obtained from finite element modelling is shown as the dotted line in Fig. 3 and indicates that the measured local critical current density values give a predicted trapped field distribution very close to that observed experimentally. The field at the edge of the sample is negative due to demagnetising effects, although no measurement was made in this region. The field at the centre is slightly lower than predicted, which can be attributed to the deviation from the Kim model at higher fields.

We conclude that the good agreement between the observed trapped field and that predicted from the finite element model using an average self-field value of $J_c$, at least at length scales larger than one millimetre, suggests that the large-scale connectivity is consistent with local values of $J_c$ determined by magnetisation measurements in $MgB_2$ bulk superconductors. As a result, further improvement in field trapping properties should be achieved via one or more of the established routes for enhancing pinning in this material [26-31]. Moreover, the results of the model indicate that caution should be exercised when comparing critical current density values calculated from field trapping measurements with those measured locally, and that the use of local self-field critical current density as a comparator can give a misleading impression of a poorly connected bulk sample.

The results reported in this investigation demonstrate clearly the potential of $MgB_2$ for high field applications at 20 K. Extensive additional experiments need to be performed, including EBSD, XRD, optical and scanning electron microscopy and magnetisation measurements, however, if the full potential of this material is to be realised, and these are the subject of an on-going study.

5. **Conclusions**

We have demonstrated the uniaxial hot pressing can be used to produce well-connected, high quality $MgB_2$ bulk samples. The critical current density determined from the trapped magnetic field at the centre of a two sample $MgB_2$ stack is consistent with measurements of the local value. This means that enhanced critical current densities, and therefore trapped fields, can be obtained more effectively by pinning enhancement rather than by further enhancing the connectivity of the $MgB_2$ bulk superconductors. The dependence of trapped field on bulk thickness means that significant enhancement of trapped field would also be expected for thicker superconducting bulk samples.

The measured trapped field of 3.14 T at 17.5 K in bulk $MgB_2$ samples fabricated by this scalable growth process is the highest reported to date for samples this small. The clear prospect of increasing trapped field by fabricating larger samples means that bulk $MgB_2$ has significant potential for practical use in high field engineering applications at 20 K.

**Figure Captions**

Figure 1. Photograph of the assembled stack of two MgB$_2$ superconducting bulk samples. The wiring for the Hall probes and the Stycast epoxy resin used to mount the sensors and join the two samples is also shown in the photograph.

Figure 2. Magnitude of the field measured by the central Hall sensor as the externally applied field was reduced from 5 T at 20 K (open squares) and 17.5 K (closed circles).

Figure 3. Profile of the trapped field measured after the removal of the external field at 20 K (open squares) and 17.5 K (closed circles). The dotted line indicates the modelled field profile expected from a local critical current density value of 8x10$^8$ A/m$^2$, observed at the centre of the sample at 17.5 K.

Figure 4. Magnitude of local critical current density determined from magnetisation measurements in a SQUID magnetometer on small specimens cut from one of the bulk MgB$_2$ samples.

**Figures**

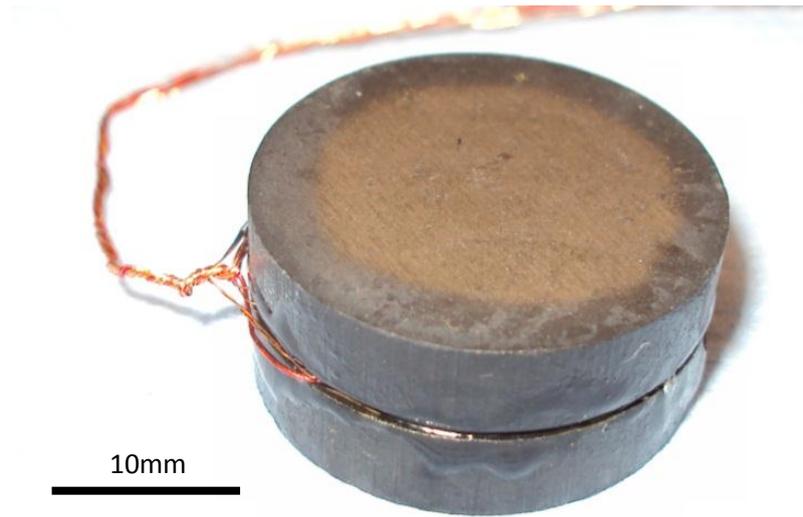

Figure 1

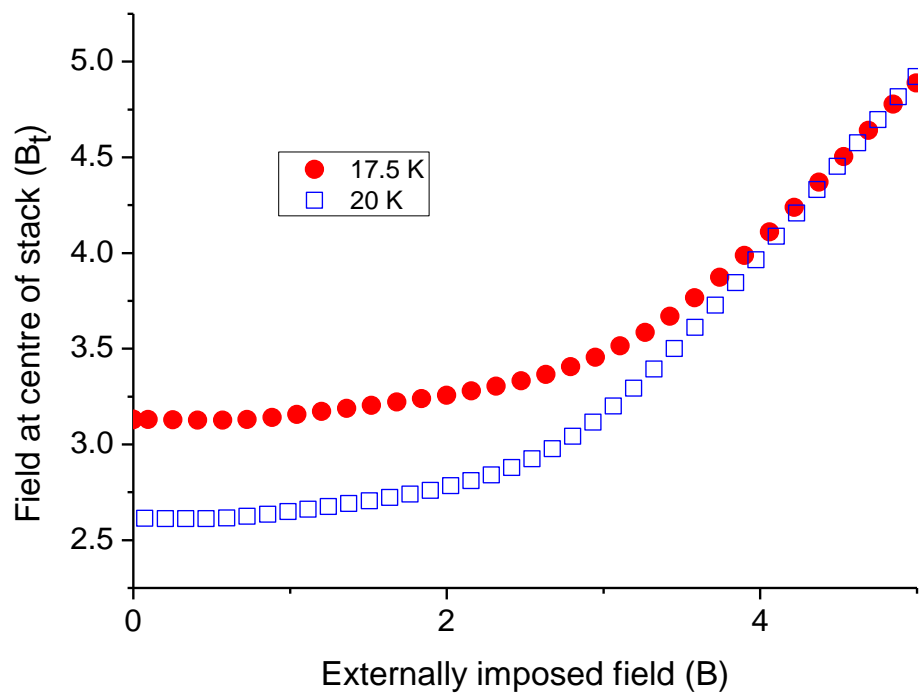

Figure 2

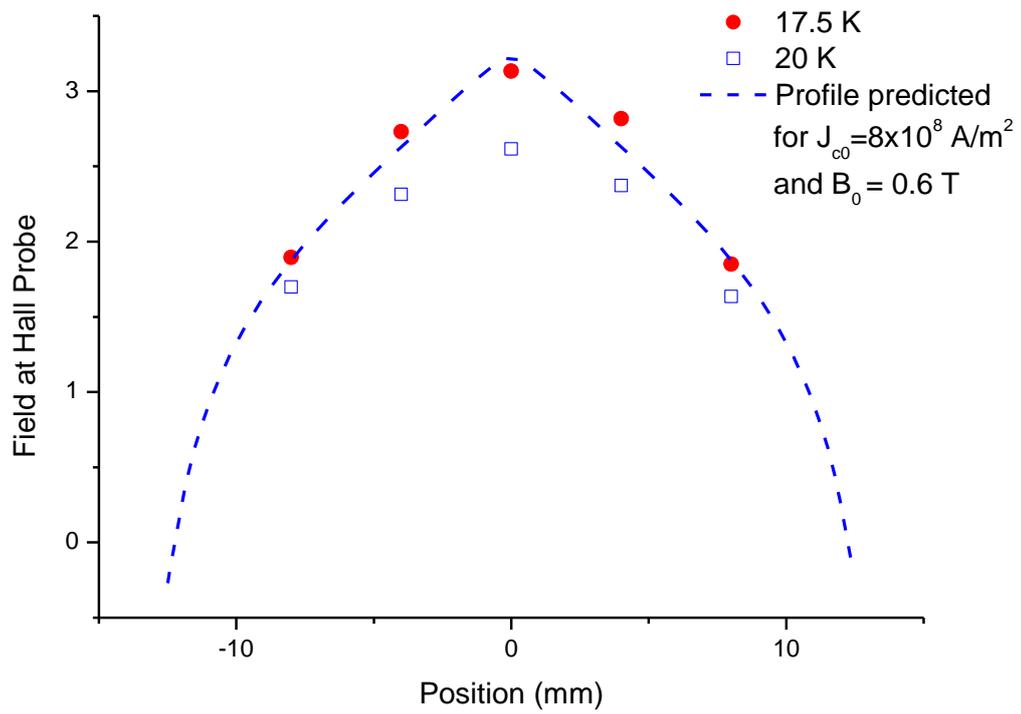

Figure 3

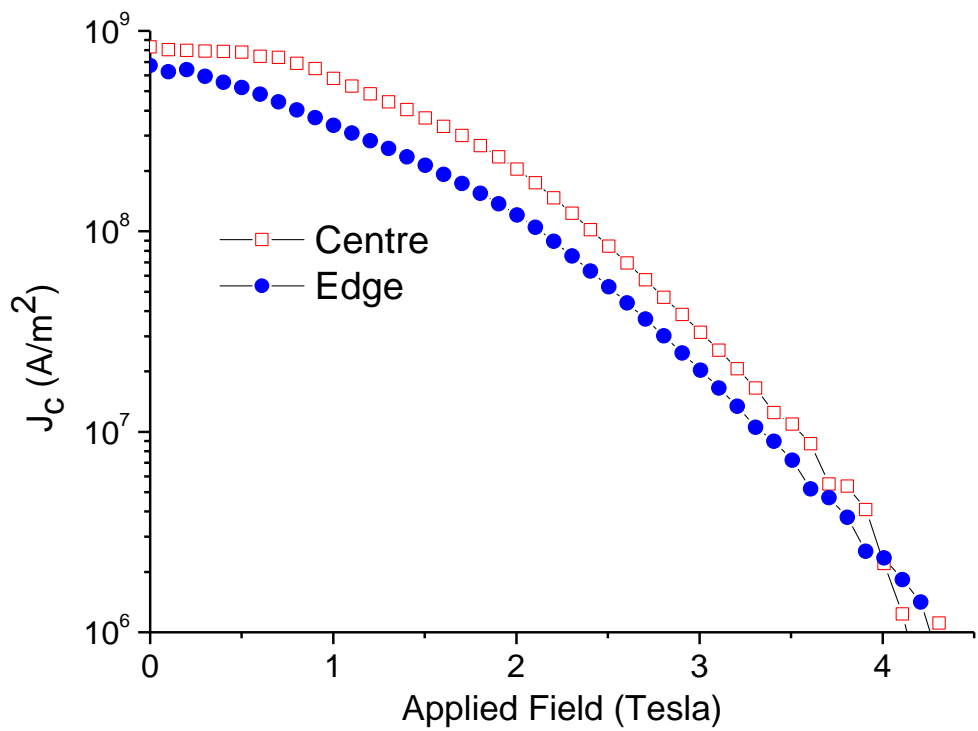

Figure 4